\documentclass{aa}  

\usepackage[varg]{txfonts}
\bibpunct{(}{)}{;}{a}{}{,}
\usepackage{graphicx}
\usepackage{multirow}
\usepackage{natbib}
\usepackage[version=4]{mhchem}
\usepackage{soul}
\usepackage[dvipsnames]{xcolor}
\usepackage{enumitem}
\usepackage{threeparttable}

\begin{document} 

\title{Ammonia snow-lines and ammonium salts desorption \thanks{The authors dedicate this paper in celebration of Stephan Schlemmer's 60th birthday on September 7, 2020. We thank all his contributions to the field of Laboratory Astrophysics.}}
\author{ F. Kruczkiewicz \inst{1,2,3} \and
         J. Vitorino \inst{2} \and
         E. Congiu \inst{2} \and
         P. Theulé \inst{1} \and
         F. Dulieu \inst{2} 
         }

\institute{Aix Marseille Univ, CNRS, CNES, LAM, Marseille, France \\
\email{franciele.kruczkiewicz@lam.fr} \and 
CY Cergy Paris Université, Observatoire de Paris, PSL University, Sorbonne Université, CNRS, LERMA, F-95000, Cergy, France
\and
Max-Planck-Institut für extraterrestrische Physik, Gießenbachstraße 1, Garching, 85748, Germany}

\date{Received --; accepted --}

\titlerunning{Ammonia snow-lines and ammonium salts desorption} 
\authorrunning{F. Kruczkiewicz et al.}

  \abstract
   {The nitrogen reservoir in planetary systems is a long standing problem. Part of the N-bearing molecules is probably incorporated into the ice bulk during the cold phases of the stellar evolution, and may be gradually released into the gas phase when the ice is heated, such as in active comets. The chemical nature of the N-reservoir should greatly influence how, when and in what form N returns to the gas phase, or is incorporated into the refractory material forming planetary bodies.}
   {We present the study the thermal desorption of two ammonium salts: ammonium formate and ammonium acetate from a gold surface and from a water ice substrate.}
   {Temperature-programmed desorption experiments and Fourier transform infrared reflection spectroscopy were conducted to investigate the desorption behavior of ammonium salts.}
   {Ammonium salts are semi-volatile species releasing neutral species as major components upon desorption, that is ammonia and the corresponding organic acid (HCOOH and \ce{CH3COOH}), at temperatures higher than the temperature of thermal desorption of water ice. Their desorption follows a first-order Wigner-Polanyi law. We find the first order kinetic parameters A = 7.7 $\pm$ 0.6 $\times$ 10$^{15}$ s$^{-1}$ and \textit{E$_{bind}$} = 68.9 $\pm$ 0.1 kJ~mol$^{-1}$ for ammonium formate and A = 3.0 $\pm$ 0.4 $\times$ 10$^{20}$ s$^{-1}$ and \textit{E$_{bind}$} = 83.0 $\pm$ 0.2 kJ~mol$^{-1}$ for ammonium acetate. The presence of a water ice substrate does not influence the desorption kinetics. Ammonia molecules locked in salts desorb as neutral molecules at temperatures much higher than previously expected that are usually attributed to refractory materials.}
   {Ammonia snow-line has a smaller radius than the water snow-line. As a result, the \ce{NH3}/\ce{H2O} ratio content in solar system bodies can be a hint as to where they formed and subsequently migrated.}

   \maketitle
%

\section{Introduction}
\label{introduction}

The abundance of nitrogen, the fifth most abundant element in the Universe, has always been a matter of debate \citep{Maiolino2019A&A}, both in the extra-galactic interstellar medium (ISM) and in the local ISM, because nitrogen has both a primary and a secondary nucleosynthetic contribution \citep{VilaCostasMNRAS93}, and because nitrogen has a significant depletion factor of 0.89~\citep{SavageSembachARAA96} indicating that a fraction of nitrogen has been incorporated into dust grains (mainly as titanium nitride, TiN). In the solar system as well, the N abundance is among the more uncertain elemental abundances \citep{Lodders10}; it has been estimated to be 7.24$\times$10$^{-5}$ with respect to hydrogen \citep{CaffauAA09,GrevesseASS10}. Nitrogen can have several carriers, depending on the phase of the ISM. In the diffuse ISM, nitrogen is mostly atomic, neutral or ionized; the column densities derived from VUV and UV absorption lines of molecular nitrogen \citep{KnauthNature04} is several orders of magnitude lower than that of atomic nitrogen \citep{NievaPrzybillaAA12}. In dense clouds, although neutral nitrogen (NI) and molecular nitrogen (N$_2$) are not observable directly, indirect observations using diazenylium (N$_2$H$^+$) \citep{MaretNature06} lead to the conclusion that atomic nitrogen is the dominant reservoir of nitrogen. 

Other N-bearing molecules such as NO, CN, NS and NH diatomic radicals, or NH$_3$, HCN, HNCO, HNC and N$_2$H$^+$ are observed with variable abundances in different sources \citep{OmontRPPh07}, NH$_3$ being the most abundant of the gas-phase carrier at a typical abundance of few 10$^{-8}$ with respect to molecular hydrogen. In pre-stellar cores it is likely that nitrogen is incorporated into the dust grain icy mantles. A recent study by \cite{Punanova2018AA} illustrates this specific property and demonstrates how N-bearing molecules can map the densest parts of star-forming regions such as filaments in the Taurus molecular cloud. The hyper-fine structure of ortho-NH$_{3}$ (1$_{0}$ -- 0$_{0}$) has been resolved for the first time in space \citep{Caselli2017AA} by observing a dark molecular cloud, confirming previous estimates of dynamics and abundances. The chemical model developed underestimates the abundance of NH$_{3}$ towards the center of the nucleus by more than an order of magnitude and overestimates its abundance by about two orders of magnitude in the outer regions \citep{HilyBlantAA10}. 
The Spitzer ice survey showed that in star-forming regions the most abundant N-bearing ices are NH$_3$, NH$_4^+$ and OCN$^-$, with typical (but with large uncertainties) abundances of about 4$\times$10$^{-6}$,  4$\times$10$^{-6}$ and 4$\times$10$^{-7}$ with respect to hydrogen nuclei, respectively \citep{BoogertARAA15}, HCN being undetected in interstellar ices. However, these three main N-carriers only account for about 10$\%$ of the overall nitrogen elemental budget \citep{ObergApJ11}. The missing N may be partially in the form of N$_2$, but the latter is almost undetectable \citep{BoogertARAA15}.
In Young Stellar Objects and outflows, where the ice is sublimated by radiative or shock heating \citep{Riaz_MNRAS_2018, Tafalla_AA_2010}, N-bearing molecules abundances increase but still cannot account for the initial atomic abundances.
In the solar system, observations by the Rosetta mission have shown that for 67/P Churyumov-Gerasimenko, like for 1P/Halley before, cometary ices are depleted in nitrogen \citep{FilacchioneSSR19}. The COSIMA instrument on board measured an average N/C of 0.035 $\pm$ 0.011, which is similar to the ratio found in the insoluble organic matter extracted from carbonaceous chondrite meteorites and in most micrometeorites and interplanetary dust particles \citep{FrayMNRAS17}. Comparing this ratio with the N/C value of 0.29 $\pm$ 0.12 in the solar system \citep{Lodders10}, we see that almost 90$\%$ of the nitrogen is missing.
Recently, one of the missing reservoirs of nitrogen may have been found in comets, as the 3.2~$\mu$m absorption band is likely to be due to ammonium salts \citep{AltweggNatA20,PochScience20}, with the spectrum of comet 67/P having the same shape of a laboratory spectrum of ammonium formate mixed
with pyrhotite grains. Ammonium salts are semi-volatile materials that desorb at higher temperatures than the more volatile NH$_3$ and H$_2$O \citep{VitiMNRAS04}, which explains why they may have been missed by the COSIMA instrument. The temperature dependence of the solid to gas phase abundance ratio of each one of the reservoirs is fundamental to address the overall N budget. The solid phase abundance of each carrier is more easily quantified after its sublimation.

Indeed, NH$_3$ is central in ice chemistry \citep{TheuleASR13}, as it bears the lone pair of
the N atom that favors the nucleophilic addition reaction with CO$_2$ giving ammonium carbamate
\ce{NH4+NH2COO-} and carbamic acid \ce{NH2COOH} \citep{BossaAA08,NoblePCCP14,PotapovApJ19}, and it acts as the base in acid-base reactions with HNCO \citep{MispelaerAA12}, HCN \citep{NobleMNRAS13}, HCOOH \citep{BergnerApJ16} and HCl, which gives the ammonium cyanate \ce{NH4+OCN-}, ammonium cyanide \ce{NH4+CN-}, ammonium formate  \ce{NH4+HCOO-} and ammonium chloride \ce{NH4+Cl-}  salts, respectively.

The NH$_3$/NH$_4^+$ hydride is an important tracer of the chemical evolution of the solar system, as it is detected in the proto-solar nebula, in the solar system bodies and in planetary atmospheres \citep{CaselliCecarelliAAR12}. This is why understanding how, and in which chemical form, it travels from the primitive molecular cloud until it is incorporated in other bodies is a key question toward understanding the complex history of the solar system formation.

Determination of the nitrogen $^{14}$N/$^{15}$N  or the hydrogen H/D isotopic ratios in different bodies of the solar system provides important information regarding the solar system's origin \citep{CaselliCecarelliAAR12}.
The $^{14}$N/$^{15}$N ratio varies greatly from body to body in the solar system \citep{HilyBlantIcarus13}: from an initial value of 441 (without corrections due to cosmic evolution) in the Solar nebula \citep{MartyEPSL12}, the Earth atmosphere, with a value of 272 \citep{MartyEPSL12}, and the solar system primitive objects, with a value of 140 in comets for all N-bearing species (HCN, CN, \ce{NH2}) \citep{RousselotApJL14} and values from 5 to 3000 in meteorites \citep{BonalGeCoA10} enriched in $^{15}$N. Similarly, the D/H ratio can vary significantly in different objects in the solar system, from carbonaceous chondrites to comets or planetary moons \citep{AlexanderScience12}, and imposes constraints on the extent of their pre-solar heritage. The D/H exchange in ice is dominated by water, by its crystallization, and is more efficient for N-bearing and O-bearing molecules rather than for C-bearing molecules \citep{FaureAA15,FaureIcarus15}.

 To make these observations meaningful, it is important to retrace the journey of the main N-carriers from the pre-solar nebula (a cold, radiation rich environment, favoring isotopic fractionation) to the solar system bodies. Although NH$_3$ and NH$_4^+$ are the main observed carriers in ice, they account for less than 10 $\%$ of the overall nitrogen elemental budget \citep{ObergApJ11}; HCN or N$_2$ are not detected in ices \citep{BoogertARAA15}. While NH$_3$ is a volatile species (it desorbs before or with water), NH$_4^+$ is a refractory species, which has been detected in cometary spectra after the water ice has sublimated \citep{AltweggNatA20,PochScience20}. Actually, the incorporation of nitrogen in the (acid) soluble or insoluble matter of chondrites,  depends on the amount and form of N-bearing molecules inherited from the proto-solar nebula ice \citep{AlexanderGeoChem17}, and the insoluble organic matter may be an important source of Earth's volatile species.

This is why reliable modeling of disk chemistry is under development \citep{WalshAA14,VisserAA18,dAngeloAA19}, with the gas/surface exchange, through accretion and desorption, being an important part of the modeling. Although such a complete modeling is clearly out of the scope of this paper, we want to show how different treatments of the ammonia desorption can lead to different conclusions. 

In its ammonium form, nitrogen is more refractory \citep{GalvezApJ10} and can be stored on grains in the inner side of the snow-line, which corresponds to the water ice mantle desorption. In the form of ammonium, the amount of nitrogen that can be incorporated into the planetesimals and later into primitive planetary atmospheres depends on its desorption mechanisms and is still an open question. Because of the temperature gradient in the protoplanetary disk, because of the migration of protoplanets in the disk and because this history can be traced through the \ce{^15N/^14N} ratio \citep{CaselliCecarelliAAR12}, the temperature dependence of the location of the different nitrogen carriers, compared to the water and methanol snow lines, is a fundamental step to understand planets formation \citep{VernazzaAJ17}. To support observations and in situ space missions, like the Rosetta mission, it is important to characterize the desorption process of species, in addition to their temperature dependent IR spectra. So far, to our knowledge, only the desorption of the ammonium cyanide salt, NH$_4^+$CN$^-$, have been quantitatively studied \citep{NobleMNRAS13}. 

In this paper, we conducted a laboratory study of the desorption of ammonium formate, \ce{NH4+HCOO-}, and ammonium acetate, \ce{NH4+CH3COO-}, salts with and without a water ice substrate. We performed temperature-programmed desorption experiments to study both the desorption mechanism and the reactivity of the two salts during the desorption process, using an experimental setup mimicking the interstellar conditions. We describe the ultra-high vacuum apparatus along with the experimental protocol and data analysis in Section~2. In Section~3 we present the results of our two complementary techniques, infrared and mass spectroscopy. We discuss the results and try to answer the following questions: what are the binding energies? When are ammonium salts formed and destroyed? Is it the ions or neutrals that mostly desorb? Is the presence of water affecting the formation and desorption of salts? In Section~4, we discuss the astrophysical implications of ammonium salts desorption, especially in the context of planetary formation.

\section{Experimental}
\label{experimental}

All experiments were conducted with the novel multi-beam ultra-high vacuum (UHV) apparatus called VENUS (VErs les NoUvelles Synthèses) based in the LERMA laboratory at the CY Cergy Paris University. VENUS has been described in details elsewhere \cite{Congiu2020} and only a brief account of the most relevant aspects of the experimental setup is given here. The experiments were carried out under UHV conditions (base pressure 2 $\times$ 10$^{-10}$ hPa at 10 K) in a stainless steel chamber. The sample holder is made of a circular copper mirror coated with gold, and mounted onto the cold head of a closed-cycle He cryostat. The sample temperature can be controlled in the 7 – 400 K range by using a regulated resistive heater clamped on the back of the sample holder. 
The ice analogs were prepared from gas condensation of molecular effusive beams aimed at close normal incidence onto the gold substrate or, in specific cases, on a slab of water ice deposited on the sample holder. The evolution of the ice during deposition was monitored by Fourier transform reflection absorption infrared spectroscopy (FT-RAIRS) using a Vertex 70 spectrometer. Once the desired ice thickness of a few layers was reached, the sample was analyzed by means of the Temperature-Programmed Desorption (TPD) technique. 

The TPD technique is used to identify the nature of desorbing species by mass and it is also a powerful method to derive binding energies of molecules adsorbed on surfaces. The method consists in increasing gradually the temperature of the surface at a constant heating rate while registering the desorption signal (a mass spectrum) as a function of sample temperature. VENUS is equipped with a Hiden 51/3F quadrupole mass spectrometer (QMS) placed 5 mm in front of the sample to monitor gas phase species in the UHV chamber. 

\begin{table}[ht]
\centering
\caption{List of experiments and corresponding ice composition, thickness and deposition temperature.}
\label{<Exp>}
\begin{tabular}{lllc}
\hline
\multirow{2}{*}{Exp.} & \multirow{2}{*}{Ice composition} & Thickness & T$_{dep}$ \\
 &  & (monolayers) & (K) \\ \hline
i & \ce{NH3} & 6.5 & 10 \\
ii & HCOOH & 4.5 & 10 \\
iii & HCOOH : \ce{NH3} & 3.7 : 4.5 & 10 \\
iv & HCOOH : \ce{NH3} & 3.7 : 4.5 & 120 \\
v & \ce{H2O} : HCOOH : \ce{NH3} & 15 : 3.7 : 4.5 & 120 \\
vi & \ce{CH3COOH} & 2.5 & 120 \\
vii & \ce{H2O} : \ce{CH3COOH} & 15 : 2.5 & 10 \\
viii & \ce{CH3COOH} : \ce{NH3} & 2.5 : 4.5 & 120 \\ 
\hline
\end{tabular}
\end{table}

The desorption characteristics of pure ices and ice mixtures (\ce{NH3}:HCOOH and \ce{NH3}:\ce{CH3COOH}) were investigated. An overview of the performed experiments is reported in Table~\ref{<Exp>}. For the mixtures (Exp. iii, iv, v, vii, viii), the proportion of ammonia was set to be 20\% higher which should favor the reaction with the acid, the same will occur in the ISM conditions since ammonia is an abundant constituent of interstellar ices which correspond to a fraction of a few percent of the ice, whereas acids are suspected to be around one percent \citep{BoogertARAA15}. The heating ramp was set to 0.2 K~s$^{-1}$ for all experiments, with each TPD starting from the deposition temperature. In the figures of this paper, the notation $\{sampleA + sampleB\}_{A : B}^T$ is used to indicate that the gases were co-deposited at temperature T and the estimated final thickness of each sample is A : B, expressed in monolayers (ML), where 1~ML = 10$^{15}$ molecules~cm$^{-2}$. As an example $\{HCOOH + NH_3\}_{3.7 : 4.5 ML}^{10K}$ means that we have co-deposited 3.7~ML of HCOOH and 4.5~ML of NH$_3$ at 10~K on the gold-coated target. The average \ce{H2O} ice coverage in ISM is supposed to be from a sub-monolayer regime up to 30~ML \citep{Potapov2020}, or even to a few hundreds according to some authors. Here the final thickness of the samples does not exceed 10~ML in the experiments without water and 25~ML in the experiment with water (Exp. v), which is a good approximation to the ISM conditions.

\section{Results}
\label{Results}

\subsection{Desorption of ammonium formate}

\subsubsection{Pure ices: HCOOH and \ce{NH3}}

\begin{figure}
   \centering
   \includegraphics[width=8.5cm]{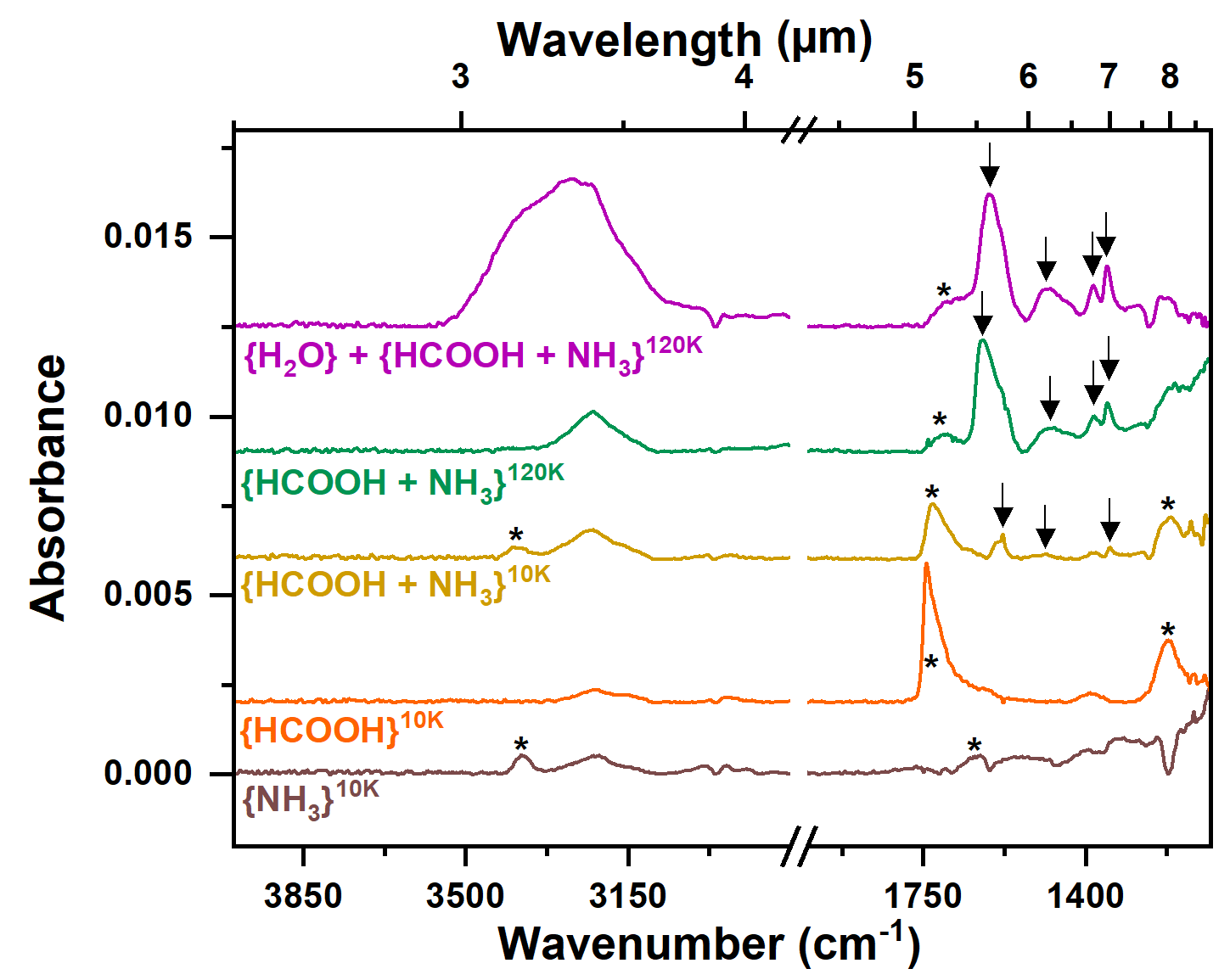}
      \caption{RAIRS spectra of experiments related to the formation of ammonium formate. The bands highlighted by stars (*) correspond to the absorption features of neutral molecules, \ce{NH3} and HCOOH. When these two species are co-deposited, vibrations of the molecular ions (\ce{NH4+} and \ce{HCOO-}) appear ($\downarrow$), suggesting the formation of ammonium formate.}
         \label{merge_FA_IR}
   \end{figure}

The IR spectra of pure \ce{NH3} and HCOOH are showed in Fig. \ref{merge_FA_IR}. The attributed absorption bands are listed in Table~\ref{<IR_FA_table>}. Both spectra were recorded at the end of the deposition of each species carried out at 10~K. The film thickness in all experiments corresponds to a few monolayers (6.5 ML for \ce{NH3} and 4.5 ML for HCOOH), for this reason the IR spectrum recorded is at the limit of signal-to-noise ratio of the spectrometer. The IR spectrum of solid \ce{NH3} is well documented in the literature \citep{DHendecourt86, Sandford1993, NobleMNRAS13, Bouilloud_MNRAS_15}, and here we can distinguish two important bands: the \ce{NH3} $\nu_3$ and $\nu_4$ modes at 3375~cm$^{-1}$ and 1641~cm$^{-1}$, respectively, indicated by stars (*) on the figure. There is also a feature around 3200~ cm$^{-1}$ attributed to crystalline water that grows on the window of the IR detector itself, located outside of the experimental chamber. This band is likely to affect the spectrum in its surroundings so extreme care was taken in analyzing features in the vicinity of 3200~ cm$^{-1}$ under our experimental conditions.

\begin{table*}[ht]
\centering
\caption{IR band assignments (cm$^{-1}$) of \ce{NH3}, HCOOH, and \ce{CH3COOH} as pure ices and as ice mixtures  \citep{DHendecourt86,Sandford1993,NobleMNRAS13, Bouilloud_MNRAS_15, BisschopAA07, HellebustJCP07, GalvezApJ10, Bahr_JCP_2006,Ito_bernstein_56,Sivaraman_APJ2013}.}
\label{<IR_FA_table>}
\begin{tabular}{lcccccccc}
\hline
\multirow{2}{*}{Mode} & \multicolumn{1}{c}{\ce{NH3}} & \multicolumn{1}{c}{\ce{HCOOH}} & \multicolumn{2}{c}{\ce{HCOOH}:\ce{NH3}} & \multicolumn{1}{c}{\ce{H2O}:\ce{HCOOH}:\ce{NH3}} & \multicolumn{2}{c}{\ce{CH3COOH}} & \multicolumn{1}{c}{\ce{CH3COOH}:\ce{NH3}} \\
 & \multicolumn{1}{c}{(i)} & \multicolumn{1}{c}{(ii)} & \multicolumn{1}{c}{(iii)} & \multicolumn{1}{c}{(iv)} & \multicolumn{1}{c}{(v)} & \multicolumn{1}{c}{(vi)} & \multicolumn{1}{c}{(vii)} & \multicolumn{1}{c}{(viii)} \\ \hline
\ce{NH3} ($\nu_4$, NH bend) & 1641 & - & - & - & - & - & - & - \\
\ce{NH3} ($\nu_3$, NH bend) & 3375 & - & 3380 & - & - & - & - & - \\
\ce{HCOOH} ($\nu$, \ce{C-O} st) & - & 1222 & 1216 & - & - & - & - & - \\
\ce{HCOOH} ($\nu$, \ce{C=O} st) & - & 1743 & 1730 & 1710 & 1697 & - & - & - \\
\ce{HCOO-} ($\nu_4$,\ce{C-O} s-st) & - & - & 1346 & 1353 & 1353 & - & - & - \\
\ce{HCOO-} ($\delta$, \ce{CH} st) & - & - & 1382 & - & 1384 & - & - & - \\
\ce{NH4+} ($\nu_4$, NH bend) & - & - & 1487 & 1475 & 1482 & - & - & 1500 \\
\ce{HCOO-} ($\nu_2$, \ce{C-O} a-st) & - & - & 1579 & 1621 & 1608 & - & - & - \\
\ce{CH3COOH} ($\nu$, \ce{C=O} st) & - & - & - & - & - & 1724 & 1716 & 1700 \\
\ce{CH3COOH} ($\nu$, \ce{C=O} a-st) & - & - & - & - & - & 1643 & 1655 & - \\
\ce{CH3COOH} ($\delta$, \ce{CH3} a-st) & - & - & - & - & - & 1413 & 1428 & - \\
\ce{CH3COOH} ($\nu$, \ce{C-O} st) & - & - & - & - & - & 1307 & 1290 & 1292 \\
\ce{CH3COO-} ($\nu$, \ce{C=O} st) & - & - & - & - & - & - & - & 1575 \\
\ce{CH3COO-} ($\nu$, \ce{C=O} a-st) & - & - & - & - & - & - & - & 1428 \\
\hline
\end{tabular}
\end{table*}

The IR properties of HCOOH were studied in details in \cite{BisschopAA07}. The formic acid spectrum in Fig.~\ref{merge_FA_IR} displays two bands at 1743 and 1222~cm$^{-1}$. The prominent peak at 1743~cm$^{-1}$ originates from the $\nu$(\ce{C=O}) stretching vibration, while the band at 1222~cm$^{-1}$ is due to the $\nu$(\ce{C-O}) stretching mode of HCOOH. \\

\begin{figure*}[ht]
   \centering
   \includegraphics[width =15 cm]{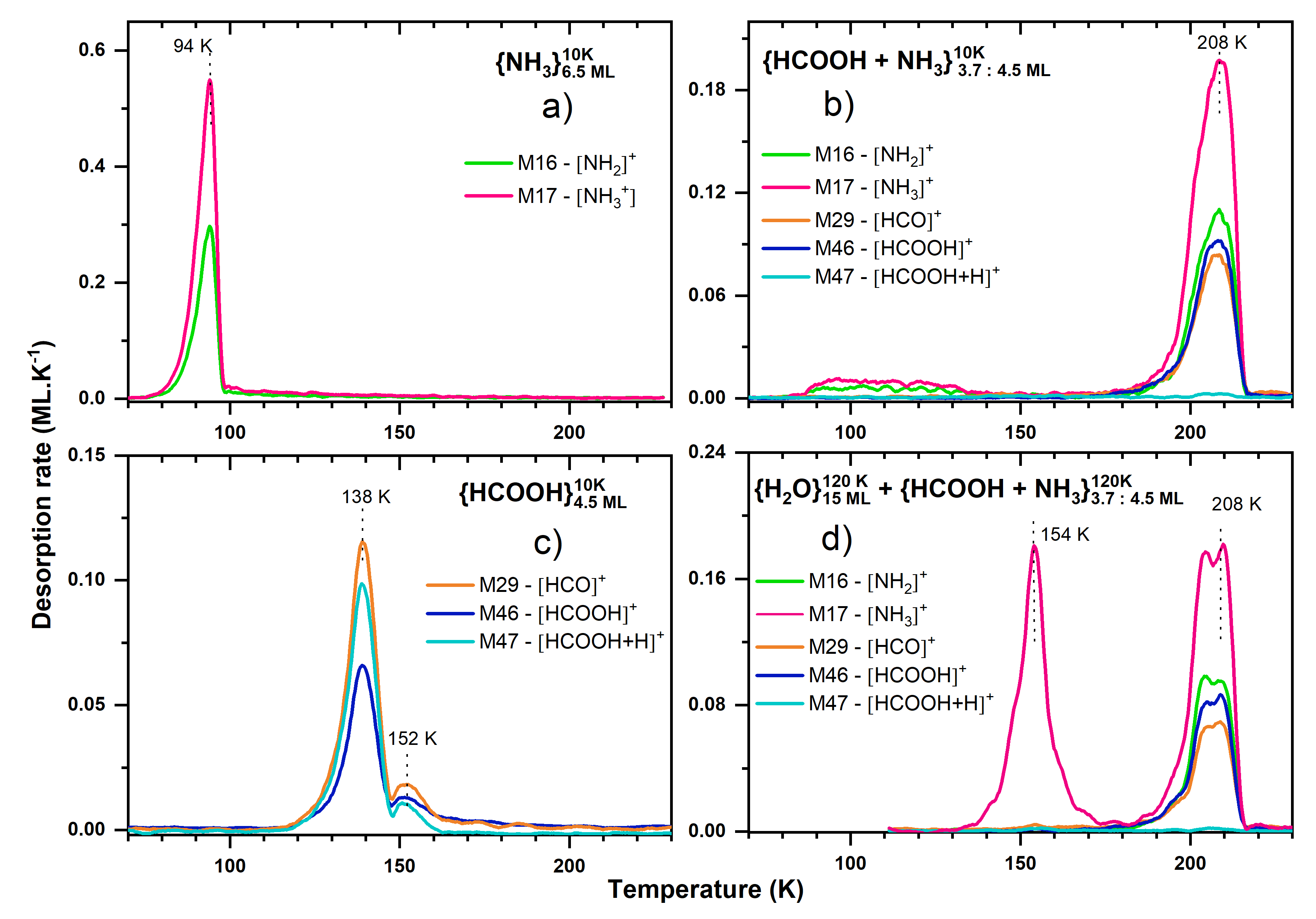}
   \caption{Left panels: TPD spectra of (a) pure \ce{NH3} and (c) pure HCOOH. Right panels: TPD spectra of mixed \ce{NH3 and HCOOH} (b) without and (d) with a substrate of water. The desorption component around 208~K traces the desorption of \ce{NH4+HCOO-} and the ratio between its fragments, similar to those of pure molecules, indicates the dissociation of the salt during the desorption process back into the components, \ce{NH3 and HCOOH}.}
   \label{merge_AF}
    \end{figure*}

Figure~\ref{merge_AF} presents the TPD profiles of the experiments involving HCOOH and \ce{NH3}. The left panels, (a) and (c), show the desorption of pure species deposited directly onto the gold substrate. In the case of pure \ce{NH3}, in addition to the parent ion NH$_{3}^{+}$, the fragment ion NH$_{2}^{+}$ was also monitored by the quadrupole mass spectrometer. The resulting TPD mass spectrum shows the desorption of the parent molecule (m/z = 17) and the fragment at m/z = 16, with both maxima of desorption peaking at around 94~K. These two mass signals always present a constant ratio of $\sim$2:1. We point out that the ratio of fragments depends on the kinetic energy of ionizing electrons, which is set to 30~eV in our spectrometer, thus lower than the typical value of 70~eV. The value of 30~eV was chosen to optimize the stability of the system and reduce fragmentation. 

In the interstellar medium, a key parameter that governs the chemical abundance in the solid and in the gas phase is the residence time of species on the dust-grain surface determined by their desorption rate. In an out-of-equilibrium treatment desorption from a surface, considering an open volume, the rate of the desorption, \(\frac{dN}{dT}\), can be described by the Wigner-Polanyi equation: 

\begin{equation} \label{WP_eq}
    -\frac{dN}{dt} = k_{des}N^n
\end{equation}

where \textit{k$_{des}$} is the desorption rate constant, \textit{N} is the number of adsorbed molecules, $n$ is the order of the reaction and the minus sign accounts for the loss of molecules from the surface. For a thermal desorption, \textit{k$_{des}$} depends on the binding energy of the specific species adsorbed, and is given by an Arrhenius law:  

\begin{equation} \label{k_des}
   -\frac{dN}{dt} = A \exp \left( - \frac{E_{bind}}{RT}\right) N^{n}.
\end{equation}

Here \textit{A} is a pre-exponential factor, E$_{bind}$ is the binding energy of the species, \textit{R} is the gas constant and \textit{T} is the surface temperature. During a TPD experiment the sample is heated at a constant rate \(\frac{dT}{dt} = \beta \) and Equation~(\ref{k_des}) can be rearranged in terms of the temperature and $\beta$ to derive the kinetic parameters, \textit{A} and E$_{bind}$, from the experimental data: 

\begin{equation} \label{PW_eq}
   -\frac{d\theta}{dT} = \frac{A}{\beta} \exp \left(\frac{-E_{bind}}{RT}\right)  
   \theta^{n}, 
\end{equation}

where \(\frac{d\theta}{dT}\) is the desorption rate (ML~K$^{-1}$), \textit{A} a pre-exponential factor in $s^{-1}$, for a first order desorption $n$~=~1, $\theta$ is the surface coverage in ML, E$_{bind}$ is the binding energy for desorption (J~mol$^{-1}$), \textit{R} is the gas constant (J~K$^{-1}$~mol$^{-1}$), and \textit{T} is the temperature of the surface (K). 

To obtain the binding energies and pre-exponential factors from the TPD curves from Equation~\ref{PW_eq}, a standard minimization of $\chi^2$ is used, which reflects the sum of the squares of the differences between the experimental profiles and calculated ones \citep{Acharyya_AA_07}. The values of desorption rate as a function of the temperature and the surface coverage were obtained directly from the experimental data. In order to work with only the binding energy as a free parameter during the fit, we used the method proposed by \cite{NobleMNRAS13}, where the pre-exponential factor value is fixed. We start assuming that the lattice vibrational frequency of the solids is $10^{13}$ s$^{-1}$. However, if the number of vibrational modes of the molecule increases, as is the case of molecules with a larger number of atoms, the lattice vibrational frequency also tends to increase and the value of the pre-exponential factor has to be optimized and increased gradually in order to reduce the $\chi^2$ value.

\begin{table*}[ht]
\caption{Kinetic parameters for the desorption of ammonium salts and their neutral components compared with available reference values. Errors arise from the spread of the fit values.}
\label{<Edes>}
\centering
\begin{threeparttable}
\begin{tabular}{lccc}
\hline
\multirow{2}{*}{Molecule} & E$_{bind}$ & A & \multirow{2}{*}{Reference} \\
 & (kJ/mol) & (s$^{-1}$) &  \\ \hline
\ce{NH3} & 25.5 $\pm$ 0.1 & 1.0 $\pm$ 0.2 $\times$ 10$^{13}$ & This work \\
& 26 $\pm$ 1 & 1.0 $\pm$ 0 $\times$ 10$^{13}$  & \cite{NobleMNRAS13} \\
HCOOH & 38.3 $\pm$ 0.1 & 1.1 $\pm$ 0.2 $\times$ 10$^{13}$ & This work \\
 & 27.8 - 47.7 & 1.0 $\times$ $10^{13}$ & \cite{Chaabouni_AA_20}$^a$ \\
 & 27.1 - 49.1 & 1.0 $\times$ $10^{13}$ & \cite{Shiozawa_JCP_15}$^a$ \\
\ce{CH3COOH} & 46.8 $\pm$ 0.2 & 1.1 $\pm$ 0.2 $\times$ 10$^{13}$ & This work \\
 & 55 $\pm$ 2 & 8 $\times$ 10$^{17\pm3}$ & \cite{Burke_2015_JPCA} \\
\ce{NH4+HCOO-} & 68.9 $\pm$ 0.1 & 7.7 $\pm$ 0.6 $\times$ 10$^{15}$ & This work \\
\ce{NH4+CH3COO-} & 83.0 $\pm$ 0.2 & 3.0 $\pm$ 0.4 $\times$ 10$^{20}$ & This work \\
\hline
\end{tabular}
\begin{tablenotes}
   \item[a] Values after the conversion with A fixed at 10$^{13}$ s$^{-1}$.
\end{tablenotes}
\end{threeparttable}
\end{table*}

Using this approach, the couple binding energy and pre-exponential factor were estimated assuming a 1st-order process (n = 1) for pure \ce{NH3}: E$_{bind}$ = 25.5 $\pm$ 0.1 kJ~mol$^{-1}$ and A = 1.0 $\pm$ 0.2 $\times$ 10$^{13}$ s$^{-1}$. These values are consistent with those previously published in the literature \citep{ULBRICHT2006, NobleMNRAS13, martin_AA_2014} and are reported with our other values in Table~\ref{<Edes>}. 

Panel c) of Figure~\ref{merge_AF} displays the TPD curve of pure formic acid. Two desorption components, at 138 and 152~K, can be distinguished in the spectra possibly because of the formation of different phases of formic acid with an amorphous and a crystalline structure. The desorption behavior of formic acid was studied previously by \cite{Bahr_JCP_05, HellebustJCP07} and recently by \cite{Chaabouni_AA_20}. Dimeric HCOOH can be present in the gas phase and remains in this form as an amorphous film during deposition being later converted into a crystalline film during annealing \citep{HellebustJCP07}. 

Formic acid dissociates into several fragments in our experiments, but here only the three major fragments are shown (Fig.~\ref{merge_AF}.c). The fragment at m/z = 46 is related to the ionized parent molecule \ce{HCOOH+}, while the fragment of higher intensity in the TPD spectra (at m/z = 29) is due to the loss of a hydroxyl group, -OH, a common pattern in carboxylic acids that gives \ce{HCO+} fragments in the case of HCOOH. The fragment at m/z = 47 can be attributed to the formation of dimeric formic acid that break up resulting in HCOOH-H$^{+}$ species \citep{Chaabouni_AA_20}. 

The binding energy and the pre-exponential factor derived using Equation~\ref{PW_eq} for formic acid are E$_{bind}$ = 38.3 $\pm$ 0.1 kJ~mol$^{-1}$ and A = 1.1 $\pm$ 0.2 $\times$ 10$^{13}$ s$^{-1}$. To compare the binding energy value with the previously published results of the kinetic parameters of formic acid, we used the method described in \cite{Chaabouni_AA_18}. Since there is a relation between A and E$_{bind}$, we can use the same pre-exponential factor (A = 10$^{13}$ s$^{-1}$) and convert the values provided by \cite{Shiozawa_JCP_15} and \cite{Chaabouni_AA_20} to compare with our results. From Table \ref{<Edes>}, we can see that the binding energy values for formic acid found in these different studies are similar, and that under our conditions of thickness and on a gold substrate, the values we find are in between the values corresponding to the HCOOH monomer and dimer. This may indicate that in our experiment, HCOOH is desorbing from the surface as a mixture of both forms of HCOOH. 

\subsubsection{Mixed ices: HCOOH + \ce{NH3}}

To form \ce{NH4+HCOO-} salt, HCOOH and \ce{NH3} gases were injected into the VENUS main chamber using two separated molecular beams, and co-deposited on the gold surface held at two different temperatures, 10 and 120~K. The flux of the \ce{NH3} beam was set to be 20\% in excess compared to the HCOOH beam, which should favor the reaction between the two species. 

Figure~\ref{merge_FA_IR} show the IR spectra of HCOOH + \ce{NH3} deposited at 10 and 120~K (displayed in yellow and green, respectively). In the experiment conducted at 10~K, it is possible to see absorption features of both species in their neutral form, highlighted in the figure with stars (*). Comparing this spectrum with the spectra of pure \ce{NH3} and HCOOH, we can distinguish the presence of NH$_3$ due to the 3380~cm$^{-1}$ absorption band, while in the low wavenumber range two features can be attributed to HCOOH: the high intensity absorption band at 1730~cm$^{-1}$ due to the \ce{C=O} stretch and the broad band at 1216~cm$^{-1}$ due to \ce{C-O} stretch. Other three bands can be distinguished in the spectrum, indicated by arrows, and are consistent with the salt bands associated with the formation of ammonium formate (\ce{NH4+HCOO-}). A broad band of low intensity around 1487~cm$^{-1}$ is attributed to the $\nu_4$ NH bending mode of \ce{NH4+} ions, and two other absorption bands are found related to \ce{HCOO-}, the $\nu_4$ \ce{C-O} asymmetric and symmetric stretch at 1579 and 1346~cm$^{-1}$ \citep{HellebustJCP07, GalvezApJ10, BergnerApJ16}. The presence of ionic features at 10~K indicates that the samples have undergone a reaction during the deposition phase. The proton-transfer between HCOOH and \ce{NH3} is an acid-base reaction that produces \ce{NH4+HCOO-} and is characterized by a low-activation energy that allows the reaction to occur at cryogenic temperatures \citep{TheuleASR13}:  

\begin{equation}\label{AF_eq}
\centering
    \ce{HCOOH + NH3 -> NH4+HCOO-}
\end{equation}

In the experiment where the mixture HCOOH + \ce{NH3} was co-deposited at 120~K the IR features related to the ion species are predominant (Fig.~\ref{merge_FA_IR}, green profile). The most intense band is due to the \ce{HCOO-} $\nu_4$ \ce{C-O} symmetrical stretch which was blue-shifted to 1621~cm$^{-1}$ compared to the spectrum deposited at 10~K. Additionally, the other salt bands present at 10~K become stronger and, in general, red-shifted: the broad band at around 1475~cm$^{-1}$ of the \ce{NH4+} $\nu_4$ NH bending mode and the $\nu_4$ \ce{C-O} symmetric stretch at 1353~cm$^{-1}$. 
We point out that in the 120~K-deposition experiment NH$_3$ molecules cannot stick and accumulate on the surface, since their desorption flux is orders of magnitude higher than their accretion flux. Therefore, only ionic forms can be kept on the surface. Despite the short residence time of \ce{NH3} (1/k$_{des}$ = 0.08 s) at 120~K we observe an efficiency of formation of salts equal to one, normalized to the HCOOH partner. This implies that the formation rate is much higher than the desorption rate. Actually, the gradual temperature increase during a TPD favors the reaction between the components producing ammonium formate \citep{HellebustJCP07, GalvezApJ10, BergnerApJ16}.

The right panels of Figure~\ref{merge_AF} display the TPD spectra of experiments involving HCOOH and \ce{NH3} mixtures deposited at 10~K without water (b) and at 120~K with water (d). The fragments displayed in the TPD curves of mixed ices are the same showed for pure species. In the experiment performed at 10~K (Fig.~\ref{merge_AF}(b)) the desorption occurs in two steps. A first desorption starts at 80~K and only fragments m/z = 16 and 17 have a relevant intensity. This peak can be attributed to the desorption of \ce{NH3} in excess, that did not react with HCOOH to produce ammonium formate during the warm-up. The second feature shows the desorption of fragments at m/z = 16, 17, 29 and 46, and occurs at around 208~K. This component desorbing at temperatures higher than the desorption of pure HCOOH and pure \ce{NH3} is attributed to the desorption of ammonium formate. The desorption energy estimated using Eq.~\ref{PW_eq} for ammonium formate is E$_{bind}$ = 68.9 $\pm$ 0.1 kJ~mol$^{-1}$ with a pre-exponential factor A = 7.7 $\pm$ 0.6 $\times$ 10$^{15}$ . Since the stoichiometry of reaction~\ref{AF_eq} is 1 : 1, the reaction yield can be estimated considering the limiting reactant HCOOH consumed during the experiment. No residual formic acid desorbs at high temperature, which suggests that all formic acid available reacted with ammonia to produce ammonium formate. Moreover, the areas of TPD peaks confirm mass conservation of the deposited species, indicating that there was no loss prior to desorption and that no change in detection efficiency of the QMS occurred.

   \begin{figure}[ht]
   \centering
   \includegraphics[width=8.5cm]{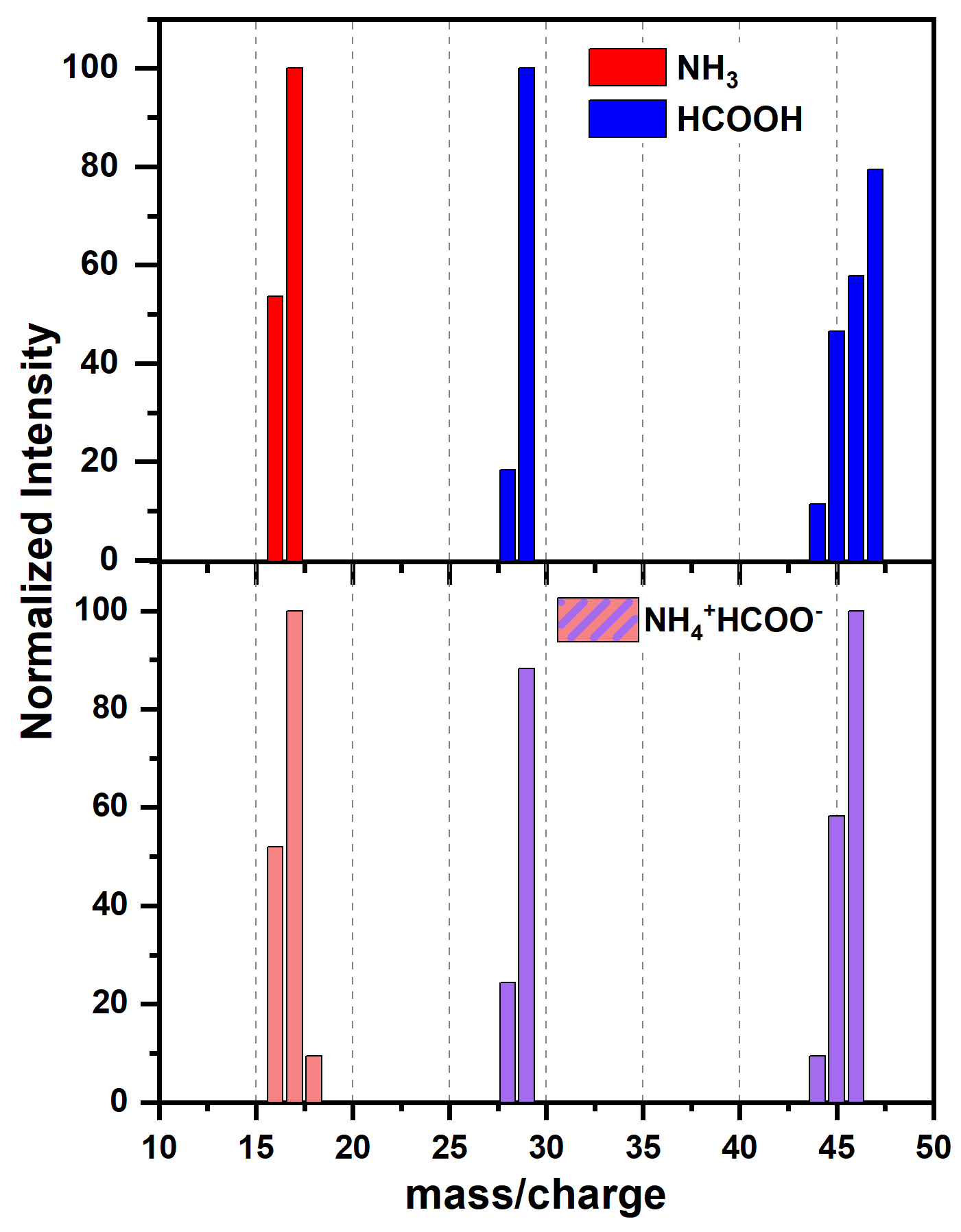}
      \caption{Mass spectra of \ce{NH3}, HCOOH and \ce{NH4+HCOO-}. Intensities are normalized to the most intense of the peaks for \ce{NH3} (m/z = 17) and HCOOH (m/z = 29). For \ce{NH4+HCOO-}, the fragments corresponding to ammonia are normalized to the most intense of the peaks (m/z = 17) and the ones from formic acid to the second peak most intense (m/z = 46).}
     \label{AF_frag}
   \end{figure}

The desorption profile of ammonium formate also indicates that the salt dissociates releasing neutral ammonia and formic acid to the gas phase at temperatures higher than the desorption temperatures of the pure species. This argument is corroborated because the same fragmentation pattern shows up both in the mass spectra of pure HCOOH and pure \ce{NH3} and in the TPD spectrum of ammonium formate (Fig.~\ref{AF_frag}). In particular, we observe the exact ratio between mass 17 and mass 16, in addition to the conservation of the total amount deposited. This would have been impossible if a substantial fraction of ammonia had desorbed in the form of \ce{NH4+}, or directly in the form of salt, which implies a different cracking pattern.

The mechanisms of sublimation of ammonium salts were investigated through \ce{NH4+Cl-} systems by \cite{Zhu_JPC-C_07}. The study found that the rate-controlling step in the dissociation of \ce{NH4+Cl-} salt is the formation and desorption of the [\ce{NH3 \bond{...} HCl}] complex that later dissociates without a distinct barrier into \ce{NH_3_{(g)}} and \ce{HCl_{(g)}}. Although in our experiment we cannot measure the formation of a [\ce{NH3 \bond{...} HCOOH}] complex, we observe that the salt dissociates and releases the neutral components into the gas phase:  

\begin{equation}\label{chem_eq_com}
\centering
    \ce{NH4+HCOO- _{(s)} <=> NH3_{(g)} + HCOOH_{(g)}}
\end{equation}

Additionally, only the fragment at m/z = 47 that is associated with the cracking pattern of HCOOH dimers is absent in the TPD spectrum of ammonium formate, suggesting that formic acid does not re-associate into its dimeric form in the gas-phase before reaching the QMS. 

\subsubsection{Influence of water ice mantles on the desorption of ammonium salts}

The experiment involving the HCOOH + \ce{NH3} mixture in the presence of water carried out at 120~K allows us to evaluate the effects of water ice mantles on the desorption of ammonium salts. The RAIRS spectrum of this experiment does not show a significant difference of the absorption features in the spectral interval between 2000~cm$^{-1}$ and 1100~cm$^{-1}$. The main difference is seen at around 3000~cm$^{-1}$ and shows the characteristic broad band of water due to the \ce{-OH} stretch. 

Panel d) of Figure~\ref{merge_AF} shows the desorption rate of ammonium formate on a water ice substrate deposited at 120~K. At around 154~K water desorption reaches its maximum as we can see looking at the ion at m/z = 17, which is 14\% of that at m/z = 18 (not shown for clarity). The second desorption peak at around 208~K exhibits the same characteristics as the mass spectra recorded in experiments performed on the bare gold target from 10 or 120~K. The water substrate does not influence the TPD profiles of \ce{NH4+HCOO-}, and the salt desorbs as a semi-volatile species after complete desorption of the water ice substrate. 

\subsection{Formation and desorption of ammonium acetate}

\subsubsection{Pure \ce{CH3COOH}}

\begin{figure}[ht]
   \centering
   \includegraphics[width=8.5cm]{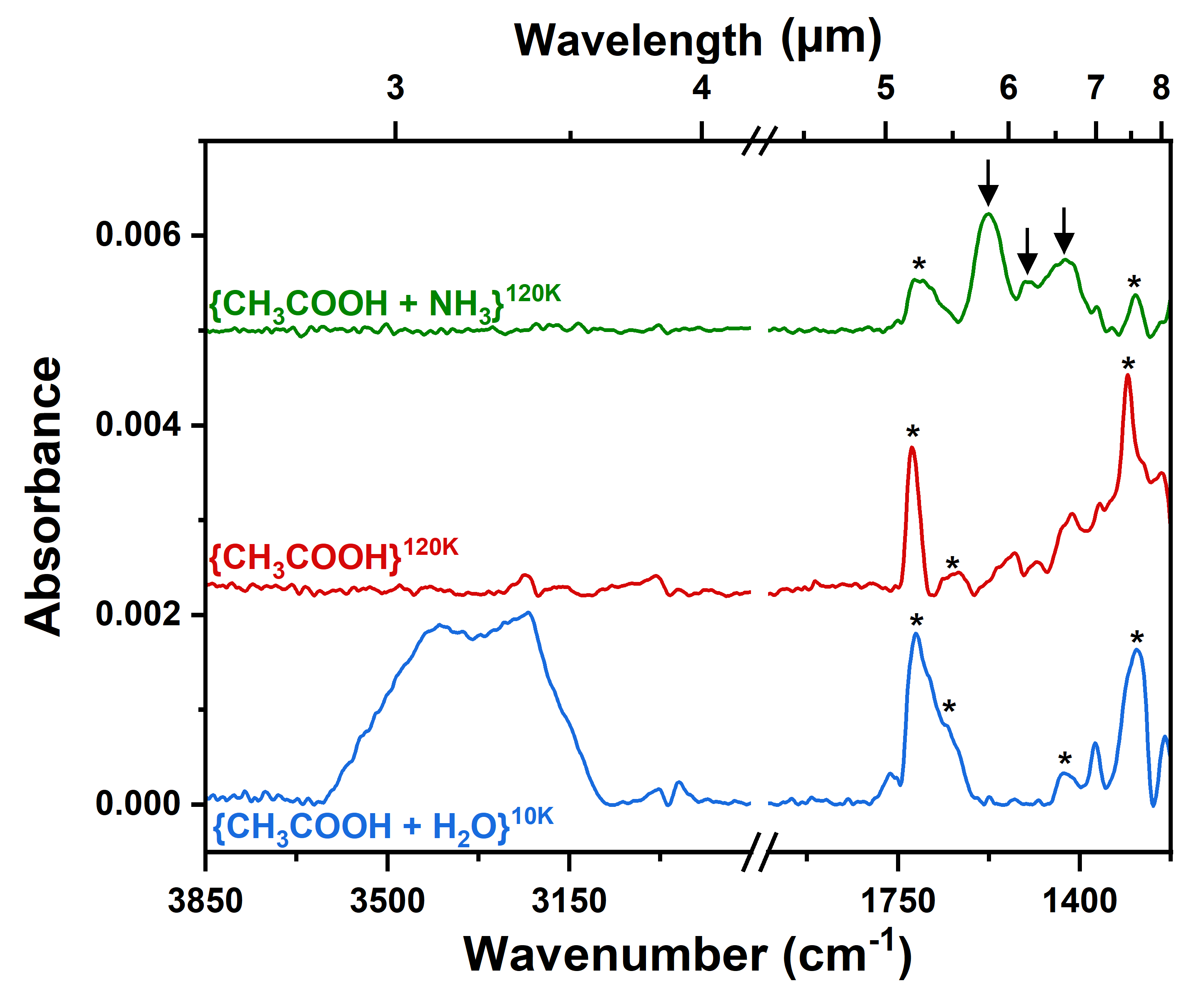}
      \caption{RAIRS spectra of experiments involving the formation of ammonium acetate. The highlighted bands using stars are related to absorption features of neutral acetic acid. When \ce{NH3 and CH3COOH} are co-deposited, the bands related to \ce{NH4+CH3COO-} appear (indicated by arrows.)}
         \label{merge_AA_IR}
   \end{figure}

The IR spectra of pure acetic acid deposited at 10 and 120~K are displayed in Figure~\ref{merge_AA_IR}. The influence of the deposition temperature and the effect of two substrates (gold and water ice) were evaluated. In the presence of water, acetic acid deposited at 10~K displays three IR bands at low frequencies assigned to the vibration of the carbonyl group. Two bands related to the $\nu$(\ce{C=O}) absorption feature at 1716~cm$^{-1}$ and 1655~cm$^{-1}$ are observed convoluted. These bands and the absorption at 1290~cm$^{-1}$ are attributed to acetic acid dimers following the values reported in previous studies \citep{Bahr_JCP_05, HellebustJCP07}. When the deposition temperature increases from 10 to 120~K the bands become narrower and resolved. The 1643~cm$^{-1}$ feature has a broad absorption and low intensity, while the 1722~cm$^{-1}$ and 1290~cm$^{-1}$ bands are sharper and predominate the spectra being the most intense bands. 

\begin{figure}[ht]
   \centering
   \includegraphics[width=8.5cm]{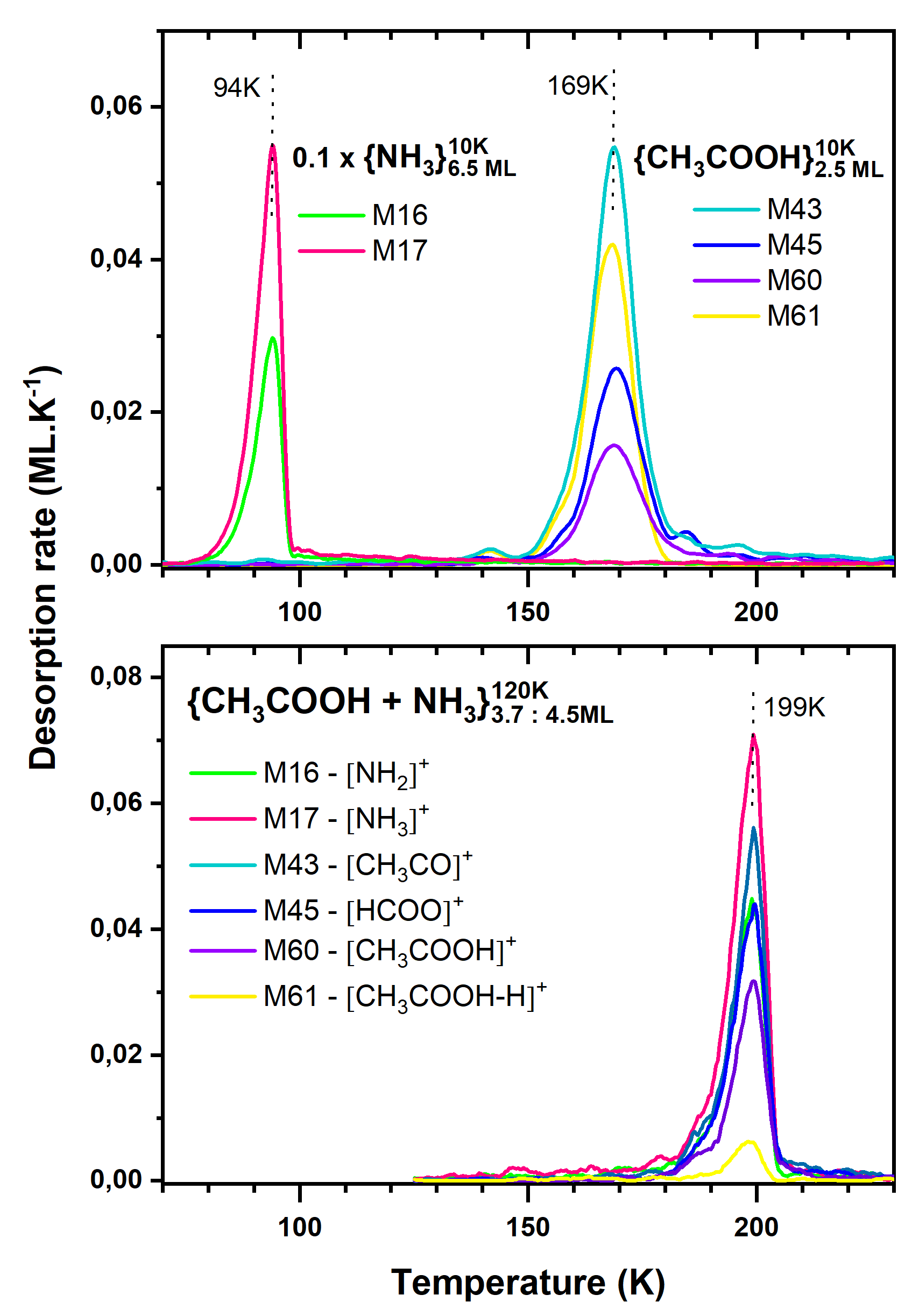}
      \caption{Top panel: TPD spectra of \ce{NH3 and CH3COOH}. Bottom panel: TPD spectrum of \ce{NH4+CH3COO-}, the sharp profile of the desorption is linked with a high pre-exponential factor and may indicate that the dissociation of the salt is at the origin of the desorption process.}
         \label{merge_AA}
   \end{figure}

The upper panel in Figure~\ref{merge_AA} shows the desorption profile of CH$_{3}$COOH and \ce{NH3} for comparison, already discussed in the previous section. Acetic acid breaks up into several species, the most intense fragment corresponding to m/z = 43, and is attributed to \ce{CH3CO+} generated after the loss of its hydroxyl group. The fragments related to the parent molecule, \ce{CH3COOH+} and the dimer, \ce{CH3COOH-H+}, are the traces for m/z = 60 and 61. 

The calculated values of binding energy and pre-ex\-ponen\-tial factor for the desorption of acetic acid are E$_{bind}$ = 46.8 $\pm$ 0.2 kJ~mol$^{-1}$ and a pre-exponential factor A = 1.1 $\pm$ 0.2 $\times$ 10$^{13}$ s$^{-1}$, that are in a good agreement with the values reported previously in the literature \citep{Burke_2015_JPCA}.

\subsubsection{Mixed ices: \ce{CH3COOH + NH3}}

The IR spectrum of the \ce{CH3COOH + NH3} mixture deposited at 120~K is showed in Figure~\ref{merge_AA_IR}. So far, to our knowledge, the IR spectrum of ammonium acetate has never been reported in the literature, so we used the IR bands of sodium acetate and methyl acetate to identify the salt features \citep{Ito_bernstein_56, Sivaraman_APJ2013}. 
The $\nu$(\ce{C=O}) asymmetrical and symmetrical stretch at 1575 and 1428~cm$^{-1}$ are attributed to \ce{CH3COO-} ions. The IR spectrum of sodium and methyl acetate do not display any prominent absorbance in the range between 1490 - 1600~cm$^{-1}$, hence we attribute the 1500~cm$^{-1}$ absorbance to the $\nu$(\ce{N-H}) bending mode of \ce{NH4+}, corresponding to the same pattern seen in the spectrum of ammonium formate.  

The bottom panel of Figure~\ref{merge_AA} presents the TPD profile of the \ce{CH3COOH + NH3} mixture deposited at 120~K. The spectra do not display traces of ammonia or acetic acid desorbing as pure species at temperatures below 180~K. The desorption peak at around 200~K is attributed to the desorption of ammonium acetate. As discussed for the TPD spectra of ammonium formate, the fragments highlighted are the same showed for pure acetic acid and ammonia and indicate that the salt dissociates releasing the base and the acid in their neutral form into the gas phase before being detected by the QMS. In the case of ammonium acetate, the fragment at m/z = 61, corresponding to one of the fragments of \ce{CH3COOH} dimers, presents a significant desorption rate, which indicates that dimeric acetic acid could form after the dissociation of the salt. 

The binding energy for ammonium acetate is estimated to be E$_{des}$ = 83.0 $\pm$ 0.2 kJ~mol$^{-1}$ with a pre-exponential factor A = 3.0 $\pm$ 0.4 $\times$ 10$^{20}$ s$^{-1}$. The high value of the pre-exponential factor is a classical trend observed when molecular complexity increases \citep{Tait+05}. Quite astonishingly, it is observed that the desorption of ammonium acetate occurs before that of ammonium formate. Usually, the larger the molecular complexes are, the greater the energy of their interaction with the surface, simply because the multiplicity of their interaction increases. This is observed, for example, in the TPD of pure formic and acetic acids, although it is a general rule of thumb (e.g., methanol has a lower binding energy than ethanol, which is lower than that of propanol). This is not what is observed for the salts, and it probably indicates that the desorption of salts occurs concurrently with dissociation into their neutral components. The binding energy would be a kind of measure of the intrinsic stability of the salt, so the ammonium formate is more stable than ammonium acetate, which is probably also related to the stronger acidity character of formic acid with respect to acetic acid, and therefore to its greater propensity to donate a proton. The ability to transfer the proton in the forward or backward reaction is translated in the  logarithm of the acidity constant (pK$_a$) of the species. This proves that what we are actually measuring is the activation energy of the endothermic backward reaction rather than the desorption energy:  

\begin{equation}\label{chem_eq_1_AA}
\centering
    \ce{NH4+CH3COO- _{(s)} <=> [\ce{NH3 \bond{...}CH3COOH}]_{(s)}} 
\end{equation}

Once the backward reaction has occurred, the neutrals, which should have desorbed long before (at 199~K t$_{res}$ = 4.9 10$^{-7}$ s for \ce{NH3} and t$_{res}$ = 1.13 ms for HCOOH), desorb very rapidly at this temperature:

\begin{equation}\label{chem_eq_2_AA}
\centering
    \ce{[NH3\bond{...}CH3COOH]_{(s)} <=> [NH3\bond{...}CH3COOH]_{(g)}}
\end{equation}
\begin{equation}\label{chem_eq_3_AA}
\centering
    \ce{[NH3\bond{...}CH3COOH]_{(g)} <=> NH3_{(g)} + CH3COOH_{(g)}}
\end{equation}

\begin{figure}[ht]
   \centering
   \includegraphics[width=8.5cm]{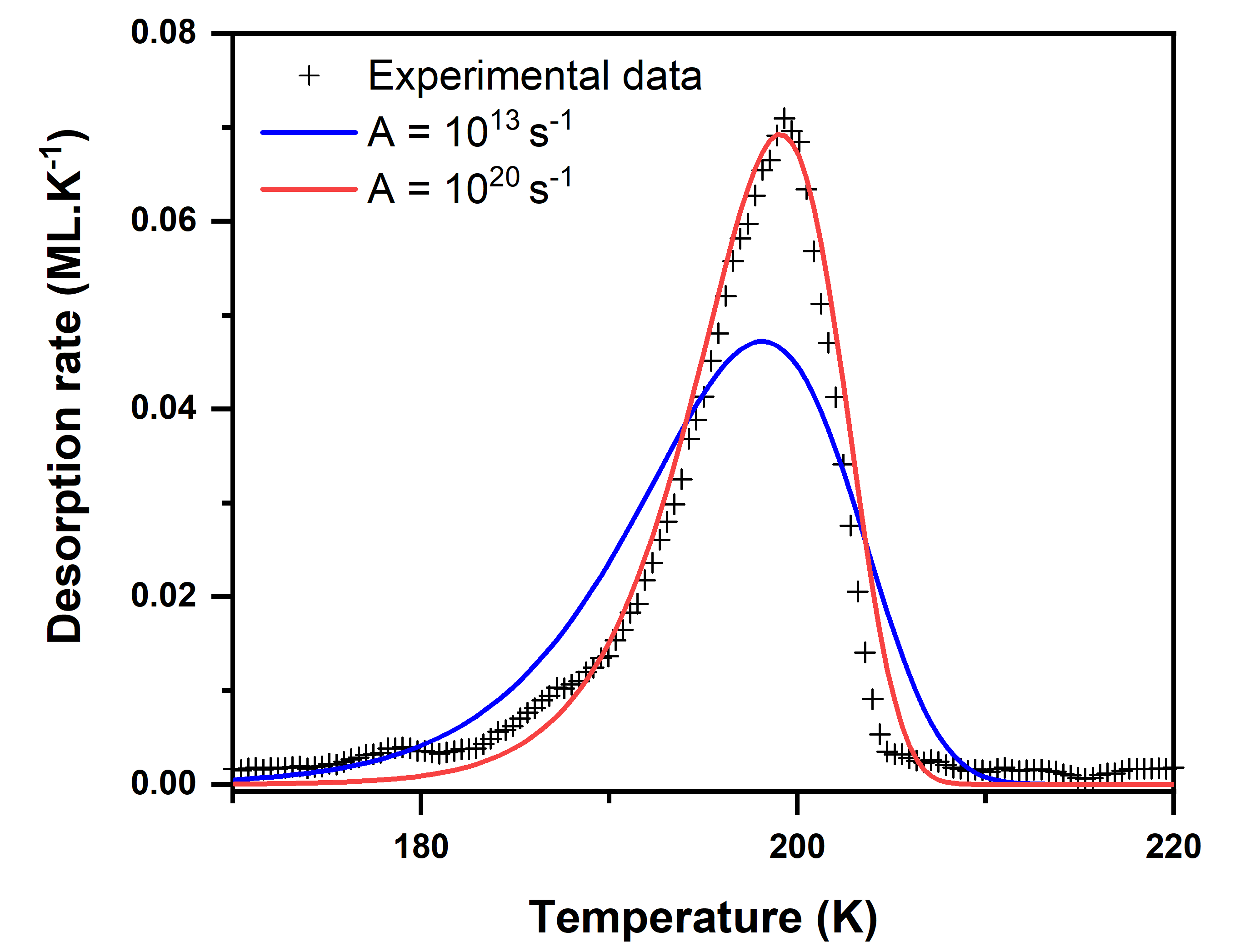}
      \caption{The TPD curve of ammonium acetate is better fitted using a high value of the pre-exponential factor, A = 10$^{20}$ s$^{-1}$, instead of A = 10$^{13}$ s$^{-1}$ used for the other molecules. The origin of this difference is probably because we are not measuring only the desorption process, but a multi-step reaction that includes the dissociation of ammonium acetate.}
      \label{fit_AA}
   \end{figure}

The pre-exponential factor hides several physical mechanisms such as a multi-step desorption composed of several elementary activated processes, or a multi-site diffusion energy, resulting in a high (and sometimes temperature-dependent) value of the factor A. Figure~\ref{fit_AA} shows the tentative fit of the acetate TPD trace using a pre-exponential factor A = 10$^{13}$ s$^{-1}$, used for the other molecules, and the best fit obtained using A = 10$^{20}$ s$^{-1}$. The fact we are measuring a (probably multi-step) chemical reaction, rather than a molecular desorption, can explain the high value of the calculated pre-exponential factor. 

\section{Discussion}
\label{Discussion}

Observations of the N-bearing carriers (the NH$_3$/NH$_4^+$ hydride, the nitrogen $^{14}$N/$^{15}$N  or the hydrogen H/D isotopic ratios in N-bearing hydrides) give information on the chemical evolution of the solar system: from the proto-solar nebula, to the solar system bodies and/or planetary atmospheres \citep{CaselliCecarelliAAR12}. Thus, it is important to retrace the chemical journey of the main N-carriers from the pre-solar nebula to the solar system bodies. NH$_3$ and NH$_4^+$ are the main observed carriers in ice, although they account for less than 10 $\%$ of the overall nitrogen elemental budget \citep{ObergApJ11}. These two species behave differently with respect to the desorption.

According to \cite{VitiMNRAS04}'s classification, NH$_3$ is a H$_2$O-like molecule, whose desorption is mostly co-desorption with water ice, and as such it can be modeled following the same pattern as the desorption of the water ice mantle. NH$_4^+$ is a semi-volatile/refractory  species (the term refractory applying to species incorporated in the dust grain by a covalent bond and sublimating at much higher temperatures than that of water ice), and the characteristics of its desorption have been measured in this paper. NH$_3$ can be easily converted into NH$_4^+$ by a set of thermal reactions involving proton transfer \citep{TheuleASR13} in the temperature range found in the warm layers of the disk or even in the disk mid-plane for a low-mass protostar, or in any region of the disk in the case of a high-mass protostar. 

\begin{figure}[ht]
\centering
\includegraphics[width=8.5cm]{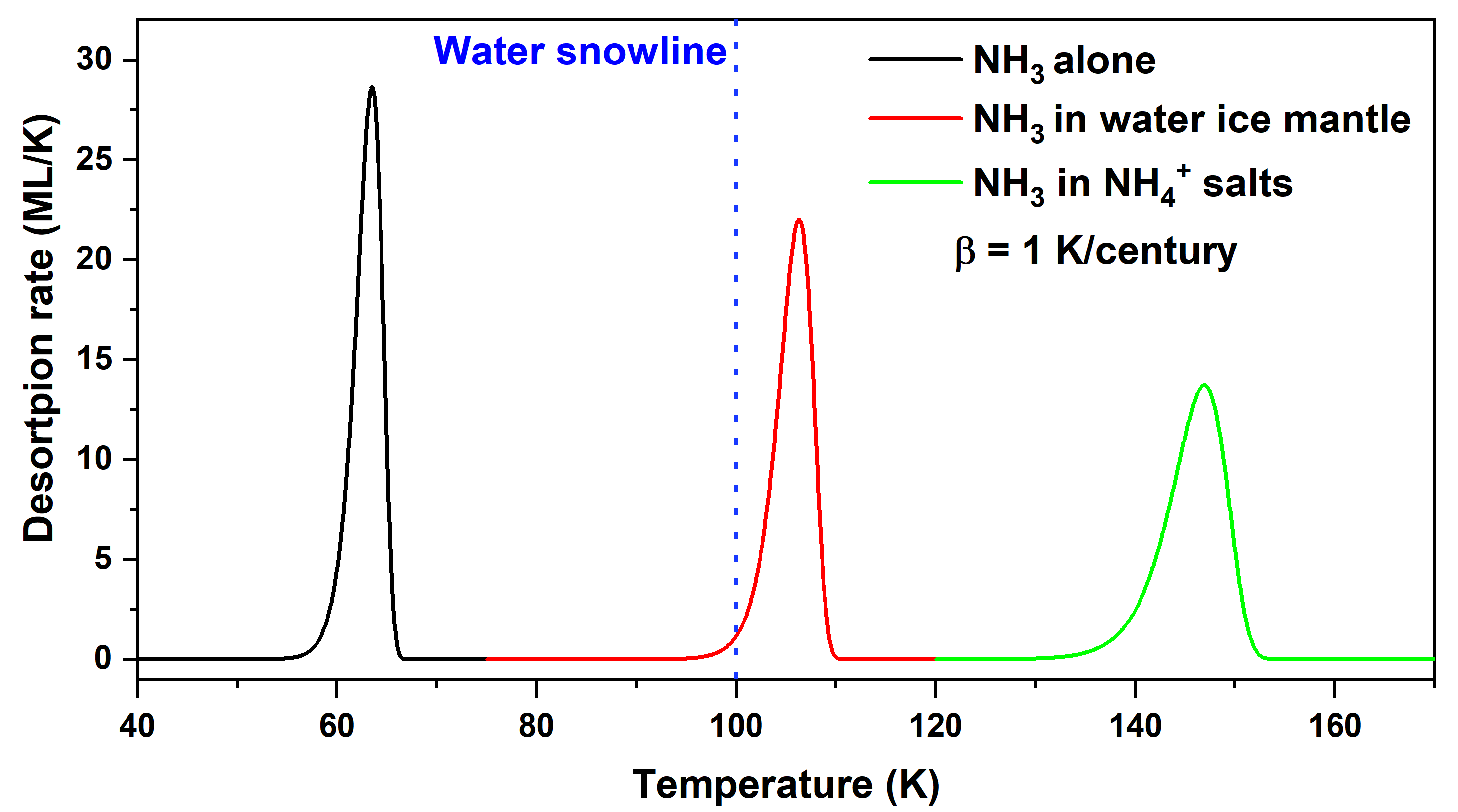}
\caption{TPD simulation showing that ammonia can desorb in more than one temperature range depending on its chemical nature and chemical environment. If ammonia is locked in ammonium salts, it desorbs at higher temperature with respect to the water snow-line. The heating rate of this simulation is $\beta$ = 1 K/century, a typical estimation for high-mass stars \citep{Viti_MNRAS_1999, VitiMNRAS04}. For low-mass stars, the heating rate would be slower and the effect of delay even more pronounced.}
\label{TPD_discussion}
\end{figure}

We will consider three scenarios: \textit{i.} the desorption of NH$_3$ (A = 1.0 $\times$ 10$^{13}$ s$^{-1}$, E = 25.5 kJ~mol$^{-1}$) desorbing from a surface as a pure species; \textit{ii.} complete co-desorption of NH$_3$/NH$_4^+$ with amorphous water ice (A = 10$^{18}$ s$^{-1}$, E = 54.5 kJ~mol$^{-1}$) that coincides with desorption from the water snow-line location in the disk; \textit{iii.} delayed desorption of NH$_4^+$ (A = 7.7 10$^{15}$ s$^{-1}$, E = 68.9 kJ.mol$^{-1}$, calculated in this work) corresponding to desorption from an "ammonium-line". We can see in Fig.~\ref{TPD_discussion} that, depending on the parameters used in the ammonia/ammonium desorption, the amount of nitrogen stored in solar system bodies located beyond the snow-line can be considerably different, as well as the implications concerning the formation of the solar system. Since ammonia is the only base, lots of species present in the ice can undergo either an acid-base reaction or a nucleophilic reaction with it \citep{TheuleASR13}. As a result, most of the NH$_3$ present in the ice should be locked in a more complex species such as an ammonium salt, which implies a delayed desorption with respect to pure NH$_3$.

Another important aspect lies in the mild temperature chemistry of semi-volatile species. While the chemistry of the parent molecules (NH$_3$, HNCO, HCOOH, CO$_2$) have been extensively studied either under conditions present in molecular clouds or during the warming resulting from a star formation process \citep{RaunierAA04,BossaAA08,BertinPCCP09,BergnerApJ16}, the chemistry of their products, the salts, have not been studied to a significant extent under conditions preponderant beyond the snow-line.  Indeed, all molecules present in the icy mantles were supposed to co-desorb with water at the snow-line or at lower temperatures if considered volatile species. The presence of semi-volatile species beyond the snow-line, at distances smaller than about 3 AU from the protostar, enables: \textit{i.} a mild (around 200 K) thermal chemistry; \textit{ii.} a dry (water-free) photo-chemistry;  \textit{iii.} a dry ion/electron-induced chemistry due to the solar winds. Since these three energetic processes are stronger beyond the snow-line, semi-volatile molecules are more likely to be included in the soluble or insoluble matter of meteorites. As an example, ammonium perchlorate salt, NH$_4^+$ClO$_4^-$ , produces hydroxylamine, NH$_2$OH, upon electron radiolysis \citep{GobiJPCA17}.

An interesting finding of this work is that most of the \ce{NH4+} is converted into NH$_3$ by a back transfer of the proton during the desorption processes, and consequently the thermal reaction NH$_4^+$HCOO$^-$ $\rightarrow$ NH$_2$CHO + H$_2$O \citep{LorinCNJPS1864} activation barrier is too high to produce formamide in detectable amounts. This is how historically formamide was produced industrially from the heating of ammonium formate \citep{LorinCNJPS1864}. In a forthcoming paper, we will address the reactivity of salts upon H-atom bombardment.

\section{Conclusion}

In this work, we presented an experimental study of the desorption of two ammonium salts, ammonium formate, \ce{NH4+HCOO-}, and ammonium acetate, \ce{NH4+CH3COO-}, in an astrophysically relevant environment. Based on our results, we conclude that: 

 \begin{enumerate}[label=({\roman*})]
 
 \item Using IR spectra of mixtures containing \ce{NH3} and an organic acid, HCOOH or \ce{CH3COOH}, we showed that ammonium salts are efficiently produced at temperatures as low as 10~K. They undergo a complete reaction with the increase in temperature, with or without the presence of water, which does not affect the desorption process under our experimental conditions.
 
\item For both salts the ammonium \ce{NH4+} transfers back its proton to the corresponding anion to release ammonia and the corresponding acid (HCOOH and
\\
\ce{CH3COOH}). No complex organic molecules (e.g., formamide, acetamide, urea) are formed during the desorption process. Ammonia is the only N-bearing species that sublimates so nitrogen atoms return to the gas phase locked in ammonia at temperatures higher than that of desorption of water ice.
 
\item Desorption of ammonium salts follows a first-order Wigner-Polanyi law. The kinetic parameters of the desorption of ammonium salts were examined. Both \ce{NH4+HCOO-} and \ce{NH4+CH3COO-} desorb following a first-order desorption with rate constant k$_{des}$= 7.7 $\pm$ 0.6 $\times$ 10$^{15} \times \exp(-68.9 \pm 0.1/RT)$ for \ce{NH4+HCOO-}, and k$_{des}$= 3.0 $\pm$ 0.4 $\times$ 10$^{20}$ $\times \exp(-83.0 \pm 0.2/RT)$ for \ce{NH4+CH3COO-}. Both salts are semi-volatile species as they desorb at temperatures higher than that of the desorption of water ice, although lower than room temperature.
 
\item In a mixture of water ice and ammonium salt, in which the salt constitutes a few percent of the total, the fraction of  \ce{NH4+HCOO-} that desorbs after the water ice mantle is 98.5$\%$ $\pm$ 0.5. This high yield shows that an important part of nitrogen is released after the water ice has completely desorbed.

\item It is fundamental to consider the chemical nature of the N-bearing molecules to correctly model their desorption in astrophysical environments. 
As we have seen in this study, the ammonium sublimation zone turns out to be closer to the protostar with respect to the water snow-line. This has important implications with regard to the nitrogen reservoir available across the protoplanetary disk.

\end{enumerate}

\begin{acknowledgements}
This project has received funding from the European Union’s Horizon 2020 research and innovation programme under the Marie Skłodowska-Curie grant agreement No 811312 for the project “Astro-Chemical Origins” (ACO). It was also supported by the Programme National “Physique et Chimie du Milieu Interstellaire” (PCMI) of CNRS/INSU with INC/INP co-funded by CEA and CNES; the DIM ACAV+, a funding program of the Region Ile de France, and by the ANR SIRC project (GrantANRSPV202448 2020-2024).
\end{acknowledgements}

\bibliographystyle{aa} 

\bibliography{bib_ammonium-salts} 

\end{document}